\documentstyle[graphicx,preprint,aps,tighten,floats,amsmath,times,epsfig]{revtex}

\def\eg{{\it e.g.}}
\def\GeV{{\;\rm GeV}}

\def\fb{{\;\rm fb}}

\def\sla#1{\ifmmode%
\setbox0=\hbox{$#1$}%
\setbox1=\hbox to\wd0{\hss$/$\hss}\else%
\setbox0=\hbox{#1}%
\setbox1=\hbox to\wd0{\hss/\hss}\fi%
#1\hskip-\wd0\box1 }

\begin{document}

\preprint{$
\begin{array}{l}
\mbox{IFT-P.042/2003}\\
\mbox{CERN-TH/2003-187}\\[-0mm]
\mbox{DESY-03-102}\\
\today\\[15mm]
\end{array}
$}

\title{Robust LHC Higgs Search in Weak Boson Fusion} 

\author{Alexandre Alves\footnote{aalves@ift.unesp.br} }
\address{Instituto de F\'{\i}sica Te\'orica, UNESP, S\~ao Paulo, Brazil}

\author{Oscar \'Eboli\footnote{eboli@fma.if.usp.br}}
\address{Instituto de F\'isica, 
Universidade de S\~ao Paulo, S\~ao Paulo - SP, Brazil}

\author{Tilman Plehn\footnote{tilman.plehn@cern.ch}}
\address{Theory Division, CERN, 
CH-1211 Geneva 23, Switzerland}

\author{David Rainwater\footnote{david.rainwater@desy.de}}
\address{DESY --T--, 
Notkestrasse 85, D-22603 Hamburg, Germany}

\maketitle

\bigskip\bigskip

\begin{abstract}
  We demonstrate that an LHC Higgs search in weak boson fusion
  production with subsequent decay to weak boson pairs is robust
  against extensions of the Standard Model or MSSM involving a large
  number of Higgs doublets. We also show that the transverse mass
  distribution provides unambiguous discrimination of a continuum
  Higgs signal from the Standard Model.
\end{abstract}

\newpage

\section{Introduction}

The search for the origin of electroweak symmetry breaking and fermion
mass generation, generally believed to be one or more scalar $SU(2)_L$
Higgs doublets, remains one of the premier tasks of high energy
physics. Fits to precision electroweak data have for some time
suggested a relatively small Higgs boson mass, in case of the Standard
Model (SM) of order $m_h \lesssim 200 \GeV$~\cite{Group:2002mc}.

The LHC will have the capability to search for physical Higgs boson(s)
over a huge mass range. In the SM case it will have multiple coverage
of search channels for any given Higgs mass, in particular for the
intermediate mass
range~\cite{Dittmar:1996ss,Rainwater:1999sd,Kauer:2000hi,WBF.taus,WBF_exp1,WBF_exp2,atlas,cms}.
This coverage relies heavily on Higgs production via weak boson fusion
(WBF). The advantage of WBF, where the scattered final-state quarks
receive significant transverse momentum and are observed in the
detector as far-forward/backward jets, is strong reduction of QCD
backgrounds due to the kinematical configuration of the colored part
of the event. In the Minimal Supersymmetric Standard Model (MSSM), all
five of the physical Higgs bosons may not be observable over all of
MSSM parameter space. However, a well-established No-Lose theorem
guarantees that at least one of the two CP-even neutral states will be
seen in WBF with subsequent decay to either tau leptons or gauge
bosons, after taking into account the LEP limits~\cite{MSSMnolose}.
For even more complex models there exists a No-Lose theorem for the
NMSSM~\cite{Ellwanger:2003jt}, which also relies on WBF Higgs
production, but includes decay modes which do not occur in simpler
models.

However, the set of possible symmetry breaking scenarios is much
larger than these models, and it is natural to consider whether the
LHC can similarly guarantee discovery for other cases. A popular way
to try to hide a Higgs sector from collider searches is via additional
fields.  These can have two effects: first, one can try to reduce the
branching fractions to observable decay modes, in analogy to invisible
Higgs decays, for example in supersymmetry. The so-called stealth
model~\cite{stealth} achieves this through the presence of additional
singlets, which couple (strongly) to the Higgs boson and lead to a
large invisible decay width. The search for these particles is tedious
for two reasons, the invisible decay and their huge width, which
vastly exceeds the experimental mass resolution and hence becomes a
limiting factor in background suppression. Nevertheless, the invisible
decay should in principle be observable at the LHC in weak boson
fusion, because the recoiling tagging jets are not sensitive to the
Higgs mass, as long as the particle is in the intermediate mass
range~\cite{invis}.

Another alternative scenario inflates the Higgs signal width via a
large number of Higgs doublets, with the measured vacuum expectation
value distributed among them~\cite{Espinosa:1998xj}. Multiple CP-even
neutral Higgs bosons exist, each with diminished coupling to weak
bosons. The WBF production rate of any one of these can be suppressed
considerably, such that by itself it would be lost in the background.
To make matters worse, each Higgs boson in general has a different
mass, forming a broad spectrum of states. Poor LHC detector resolution
may smear the individual peaks out to a smooth continuum.  These
models lead to the speculation that Higgs sector may be unidentifiable
at the LHC, even though high-energy $WW$ scattering would clearly
observe unitarity~\cite{Gunion:2001pg}.

To decide whether there really is a hole in the LHC Higgs boson
discovery potential, we investigate these continuum models in detail.
Our analysis of WBF production of such a continuum of Higgs bosons,
followed by decay to dileptons ($e^\pm \mu^\mp$, $e^\pm e^\mp$,
$\mu^\pm\mu^\mp$) via a pair of $W$ bosons confirms the robustness of
the WBF Higgs searches in the case of non--standard Higgs sectors.
The signal can easily be observed, and typically distinguished from a
SM Higgs sector with only modest integrated luminosity, ${\cal
  O}(100)\fb^{-1}$, via study of the transverse mass distribution.

\section{Multiple Higgs doublets}
\label{sec:model}

The continuum model~\cite{Espinosa:1998xj} is merely a generalization
of the SM to an arbitrary number of $SU(2)_L$ Higgs
doublets~\footnote{Additional singlets may also appear, but we
  disallow the presence of Higgs triplets which violate the custodial
  $SU(2)$ symmetry.}. The vacuum expectation values $v_i=C_i v$, are
bound by the measurement
\begin{equation}
\sum_i v^2_i = v^2 \, \sum_i C_i^2 \geq \, v^2 \, \equiv (246 \GeV)^2
\end{equation}
with the equal sign $\sum C_i^2 = 1$ in the case of only singlet and
doublet Higgs fields which we consider in this letter. Each doublet
gives rise to one CP-even neutral Higgs boson $h_i$. There are in
addition a large number of charged and CP-odd neutral Higgs bosons
which can all be sufficiently heavy as to be unobservable. The
question arises: given that none of the heavy states could be found,
can LHC discover any of the $h_i$, and if so can it distinguish them
from the SM Higgs boson?

The gauge couplings of the Higgs boson to weak bosons is proportional
to $gv$, therefore it is modified for the CP-even Higgs boson of each
doublet by the single factor $C_i$. The partial width of a Higgs to
weak bosons is then $C_i^2 \; \Gamma^{\rm SM}_{W,Z}$. A further
constraint from electroweak precision data is~\cite{Gunion:2001pg}
\begin{equation}
\label{constraint}
\sum_i C^2_i m^2_{h_i} \, = \, \langle M^2 \rangle \, 
\lesssim (200 \GeV)^2 
\end{equation}
which also holds for any general supersymmetric Higgs
sector~\cite{Espinosa:1998re}.

Each SM fermion gets its mass in small pieces from the many Higgs
fields: $m_f = \sum_i Y^i_f v_i$. In general, these Yukawa couplings
need not be equal. In the minimal model, which assumes that ultimately
the flavor sector comes from some universal condition, each $SU(2)_L$
Higgs doublet experiences the same set of Yukawa couplings $m_f = Y_f
\sum_i v_i$. Matching the universal Yukawa coupling $Y_f$ with its SM
counterpart we are left with $Y_f^{\rm SM} = Y_f \sum_i C_i$. Hence,
the Higgs' partial widths to fermions acquire scaling factors
$C_i^2/(\sum_i C_i)^{-2}$. The Higgs decay properties are particularly
close to the SM properties in the limit of $N$ Higgs fields with
identical vacuum expectation values $v_i=v/N$. Each partial width to
$W,Z$ bosons becomes $\Gamma_{W,Z} = \Gamma^{\rm SM}_{W,Z}/N^2$. The
partial widths to fermions are scaled by the same factor $\Gamma_f =
\Gamma^{\rm SM}_f/N^2$.  Thus, to leading order all branching
fractions for one of the many Higgs fields are identical to the SM
values for the same Higgs mass; only the total width is suppressed by
a factor $N^{-2}$.

The limiting case of a large number of Higgs fields is a Higgs
continuum~\cite{Espinosa:1998re}. In this case, the distribution of
vacuum expectation values becomes a continuous quantity, $C(m_h)$,
expressed as a function of the physical Higgs boson mass. The sum
rules in this case become:
\begin{equation}
\label{contin}
\int dm_h \, C(m_h)^2 \, = \, 1 \qquad \qquad
\int dm_h \, C(m_h)^2 m^2_h  \, \lesssim \, (200 \GeV)^2 \, .
\end{equation}

As pointed out in Ref.~\cite{Espinosa:1998re}, the weak production
cross section for any point in the physical Higgs continuum is
minimized when all the couplings to gauge bosons are set equal. This
is precisely the continuum limit of the model with $N$ equal values
$v_i$, which leaves the Higgs branching fractions unaffected to
leading order. Taking the direct LEP constraints~\cite{:2003sz}
together with the precision data, we find a scenario which is
maximally likely to escape detection: $C(m_h)=C_0$ for 70 GeV $\le
m_h\le 300\GeV$ and $C(m_h)=0$ elsewhere. We adopt this case to
illustrate our analysis, then slightly vary this assumption in
Sec.~\ref{sec:disc}.

\section{The Weak Boson Fusion Process}
\label{sec:calc}

We simulate $pp$ collisions at the LHC, $\protect\sqrt{s}=14$~TeV, for
the final state $\ell_1^+\ell_2^-jj\sla{p}_T$, calculating all signal
and background cross sections with full tree level matrix elements for
the contributing subprocesses. $\ell_1,\ell_2$ are any combination of
$e$ and $\mu$, which are easily identified by the detectors with high
efficiency. We employ CTEQ6L1 parton distribution
functions~\cite{Pumplin:2002vw} throughout. Unless otherwise noted the
factorization scale is chosen as $\mu_f =$ min($p_T$) of the defined
jets. The signal is calculated with matrix elements constructed by
Madgraph~\cite{Stelzer:1994ta}, including exact matrix elements for
the decay $H\to W^+W^-\to \ell_1^+\ell_2^-\nu_1\bar\nu_2$ to maintain
the decay helicity correlations~\cite{Dittmar:1996ss} on which the
analysis rests.

The backgrounds consist of both QCD and EW processes leading to two
far forward/backward jets, with a central pair of oppositely-charged
leptons $e$ or $\mu$, and large missing transverse energy. Although
QCD processes with the same number of final state partons have much
larger cross sections than the counterpart EW processes, in the region
of phase space with two tagging jets the EW contribution can easily be
of the same size as the QCD component. We thus consider the QCD
processes $t\bar{t}$~+jets, $W^+W^-jj$ and $\tau^+\tau^-jj$, and the
EW processes $W^+W^-jj$ and $\tau^+\tau^-jj$~\cite{Rainwater:1999sd},
including correction factors for off-shell top quark
effects~\cite{Kauer:2000hi,OS-tops}.

We also include here for the first time as a background $W$-fusion
single-top production~\cite{Willenbrock:cr}, $pp\to tbj$, where the
light jet and hard-scattered $b$ jet appear as the far
forward/backward tagging jets, the top quark decays semi-leptonically,
and its daughter $b$ quark also decays semi-leptonically.  We use
exact matrix elements for the $b$ decay to both charm and up quarks.

\subsection{Basic kinematic cuts and jet selection}
\label{sec:cuts}

The characteristics of WBF Higgs boson production are a pair of very
far forward/backward tagging jets with significant transverse momentum
and large invariant mass between them. Furthermore, the Higgs boson is
produced centrally, and the decay products will therefore typically
lie between the tagging jets, independent of the Higgs boson decay
mode. Since the only modification to our signal here is a spectrum of
Higgs boson masses rather than a single resonance, we employ the same
optimized jet cuts as in the experimental simulation presented of
Ref.~\cite{WBF_exp1}. These do not vary significantly from those of
the original study~\cite{Rainwater:1999sd}, but do include more
current understanding of detector and trigger requirements:
\begin{alignat}{7}
\label{eq:jbasic}
&   p_{T_j} > 40,20 \GeV \; , \qquad 
&& |\eta_j| < 4.9 \; , \qquad
&& \triangle R_{jj} > 0.4 \; , \qquad \notag \\
&   \eta_{j,{\rm min}} < \eta_\ell < \eta_{j,{\rm max}} \; , \qquad
&& \eta_{j_1} \cdot \eta_{j_2} < 0 \; , \qquad
&& |\eta_{j_1}-\eta_{j_2}| > 3.8 \; , \qquad
&& m_{jj} > 550 \GeV  \; .
\end{alignat}

The top pair background will frequently contain an extra central $b$
jet with $p_{T_b} > 20 \GeV$ and $|\eta_b| < 2.5$. We veto these in
the same manner as in Ref.~\cite{WBF_exp1}. 

Although we are now examining a continuum of Higgs boson masses, they
decay in the same manner as in the previous studies. The lepton cuts
used therein take into account the detector observability and the
angular correlations of the decay. Since neither change fundamentally,
we also use the same lepton cuts as in
Ref.~\cite{WBF_exp1}:~\footnote{A separate parton level
  study~\cite{Kauer:2000hi} modified these cuts to optimize for a
  lighter Higgs boson, but in this analysis we focus on large Higgs
  masses, as will become obvious later. We are use the cuts of
  Ref.~\cite{Kauer:2000hi} for the same--flavor lepton channel.}
\begin{alignat}{7}
&  p_{T_\ell} > 15,20 \GeV \; , \qquad
&& |\eta_{\ell}| \leq 2.5 \; , \qquad 
&& \triangle R_{j\ell} > 0.4 \; , \qquad 
&& p_{T_\ell} < 120 \GeV \; , \qquad \notag \\
&  \cos \theta_{e\mu} > 0.2 \; , \qquad
&& \phi_{e\mu} < 1.05 \; , \qquad
&& \triangle R_{e\mu} < 1.8 \; , \qquad
&& m_{e\mu} < 85 \GeV \; .
\label{eq:lbasic}
\end{alignat} 
Along with the $\triangle R_{j\ell}$ cut is the lepton isolation
criteria for the semileptonic $b$ decay in $t\bar{t}j$ events: the
hadronic remnant must have $p_T<3$~GeV if it lies within a cone of
$\triangle R<0.4$ of the lepton.

The real tau backgrounds are large, especially the QCD component, but
are reducible by reconstructing these
taus~\cite{Ellis:1987xu,WBF.taus}. In the collinear decay
approximation, the fraction of tau energy that each charged lepton
takes with it in the decay ($x_\tau$) is solved for the actual charged
lepton momenta and the missing energy in the transverse directions.
Events with real tau pairs typically have $\sla{\vec{p}}_T$ lying
between the charged lepton flight directions; leptonic $WW$ events
typically do not, and will most often give a fake--tau reconstruction
with negative $x_\tau$ values. We therefore reject events with two
positive $x_\tau$ values and an invariant tau pair mass of $m_Z \pm 25
\GeV$~\cite{Rainwater:1999sd,Kauer:2000hi}.

QCD $b\bar{b}jj$ events with dual semileptonic $b$ decays constitutes
a very large background to EW $WWjj$ events~\cite{Kauer:2000hi}.
However, they typically give a small transverse momentum for the
reconstructed Higgs boson, because the leptons do not pass the
isolation cut from the charm quark unless the parent $b$ quark is
soft. We therefore employ dual two-dimensional
cuts~\cite{Kauer:2000hi} which suppress the $b\bar{b}jj$ background by
more than two orders of magnitude:
\begin{equation}
\label{eq:2D}
\triangle\phi_{\ell\ell,{\rm miss}} + 1.5 \, p_{T_H} > 180 
\; , \qquad
12 \, \triangle\phi_{\ell\ell,{\rm miss}} + p_{T_H} > 360 \; .
\end{equation}
We then reduce QCD $b\bar{b}jj$ events to an insignificant level by
imposing a conditional cut on the transverse
momentum~\cite{Kauer:2000hi}:
\begin{equation}
\label{eq:cond}
\sla p_T > 20 \; {\rm GeV} \qquad
\hbox{provided} \qquad p_{T_H}<50 \; {\rm GeV} \; ,
\end{equation}
where 
$p_{T_H}=|\vec{p}_{T}(\ell^+)+\vec{p}_T(\ell^{\prime -})+\sla{\vec{p}}_T|$.  
Both cuts result in only trivial signal rejection.

In the case of same lepton flavors, low mass $\ell^+ \ell^-$ pairs
originating from $\gamma^\star \to \ell^+ \ell^-$ exhibit a large
cross section and are suppressed by requiring that the dilepton
invariant mass is larger than $10 \GeV$~\cite{Kauer:2000hi}.  To
reduce the background arising from $\ell^+ \ell^- j j$, where the
missing transverse momentum is generated by detector effects, we
further require the missing transverse momentum to be larger than $30
\GeV$~\cite{Kauer:2000hi}. In brief, we impose the additional cuts on
the same flavor final state:
\begin{equation}
  m_{\ell\ell} > 10 \GeV   \qquad \hbox{and} \qquad
  \sla{p}_T > 30 \GeV \; .
\end{equation}

Finally, we note that the single-top background is dominated by the
semileptonic $b$ decay $b\to u\ell\nu$, rather than $c\ell\nu$, although 
the CKM matrix element is larger in the latter case.  Because
the kinematics of the lighter quark decay allows more momentum of the
light quark transverse to the $b$ flight direction, it can more often
lie outside the lepton isolation cone.

\subsection{Discovery Potential of the LHC}
\label{sec:disc}

\begin{figure}
\begin{center}
\epsfig{file=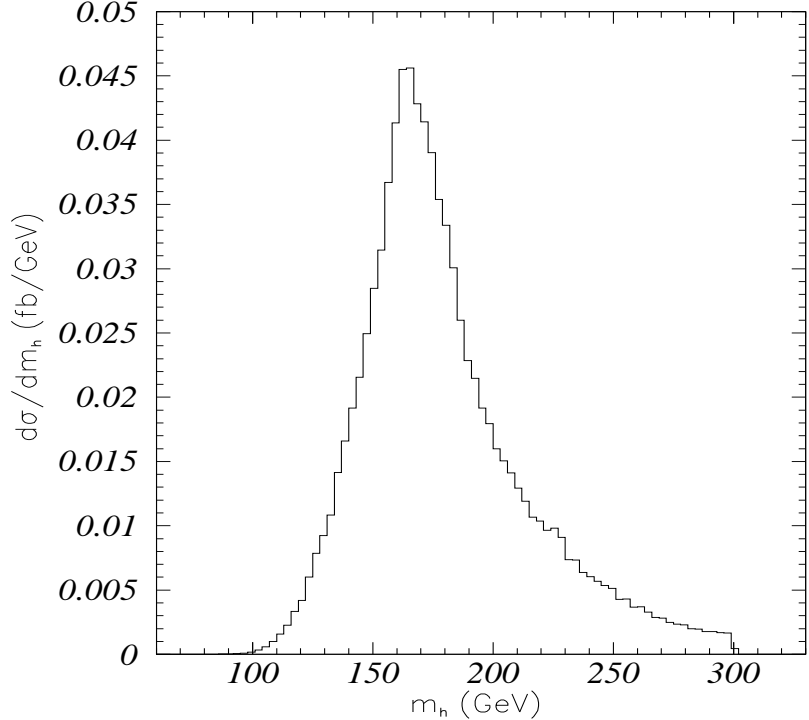,width=0.75\linewidth,height=0.65\linewidth}
\end{center}
\vspace{-2mm}
\caption{Distribution of the signal cross section after all cuts, 
as a function of the continuum Higgs mass, for the case 
$70\leq m_h\leq 300\GeV$, as discussed in the text.}
\label{fig:1}
\end{figure}

The size of the signal which could be observed at the LHC in the
channel described above depends critically on the continuum mass
window.  We know that for flat $C(m_h)=C_0$ the branching fraction to
$WW$ is reduced for small Higgs masses. In other words, the continuum
signal at low masses will be suppressed by the $WW$ branching
fraction, despite the larger Higgs production cross section.  Large
continuum masses will be production phase space suppressed, the $ZZ$
branching fraction becomes significant, and the selection cuts are not
optimized for $m_h \gtrsim 200 \GeV$.  Hence, even a model with a wide
range of continuum masses will still leave the event sample dominated
by Higgs masses around the $WW$ threshold. In Fig.~\ref{fig:1} we show
the contribution from a continuum Higgs sector, $70\leq m_h\leq
300$~GeV, to the total $W^+W^-jj$ cross section.  As expected, the
bulk of the signal events originate from Higgs masses around $2M_W$.

As the final state events contain two unobservable neutrinos and are
thus not fully reconstructible, we cannot directly observe the $m_h$
distribution in Fig.~\ref{fig:1}.  However, the general behavior can
be read off the measurable $WW$ transverse mass
distribution~\cite{Rainwater:1999sd},
\begin{equation}
m^2_T \equiv
m^2_{T,WW} = \left( \sqrt{ \vec{p}_{\ell\ell,T}^{\; 2} + m_{\ell\ell}^2 }
                   +\sqrt{ \vec{p}_{\nu\nu,T}^{\; 2} + m_{\nu\nu}^2 } \;
             \right)^2
           - \left( \, \vec{p}_{\ell\ell,T} + \vec{p}_{\nu\nu,T} \,
             \right)^2 \; ,
\end{equation}
which for a single resonance exhibits a sharp edge near the invariant
$WW$ mass $m_{WW}$.  The neutrino pair transverse momentum is
determined as the missing transverse momentum $\vec{p}_{\nu\nu,T} =
\sla{\vec{p}}_T$.  The invariant mass of the neutrino pair cannot be
observed, so we replace it with the lepton pair invariant mass
$m_{\nu\nu} \sim m_{\ell\ell}$~\cite{Rainwater:1999sd}, an
approximation which is exact for $m_h = 2M_W$, and very good over the
intermediate Higgs mass range\footnote{In the limit
  $m_{\ell\ell},m_{\nu\nu} \to 0$ the $WW$ transverse mass can be
  written as $m_{T,WW}^2=2 p_{\ell\ell,T} \, \sla{p}_T \, (1-\cos
  \Delta\Phi)$. This alternative definition can be used to distinguish
  signal and background, but does not work particularly well in
  distinguishing continuum Higgs sectors from the SM.}. On average the
two kinds of $W$ decay fermions will give the same distributions, even
though on an event-by-events basis this approximation violates the
condition $m_T \leq m_{WW}$. Detector effects, mostly mismeasurement
of the missing transverse momentum due to the presence of the hadronic
forward tagging jets, will dull the edge in $m_T$ somewhat, but the
distribution remains extremely important~\cite{WBF_exp1,WBF_exp2}.

\begin{figure}[t]
\begin{center}
\epsfig{file=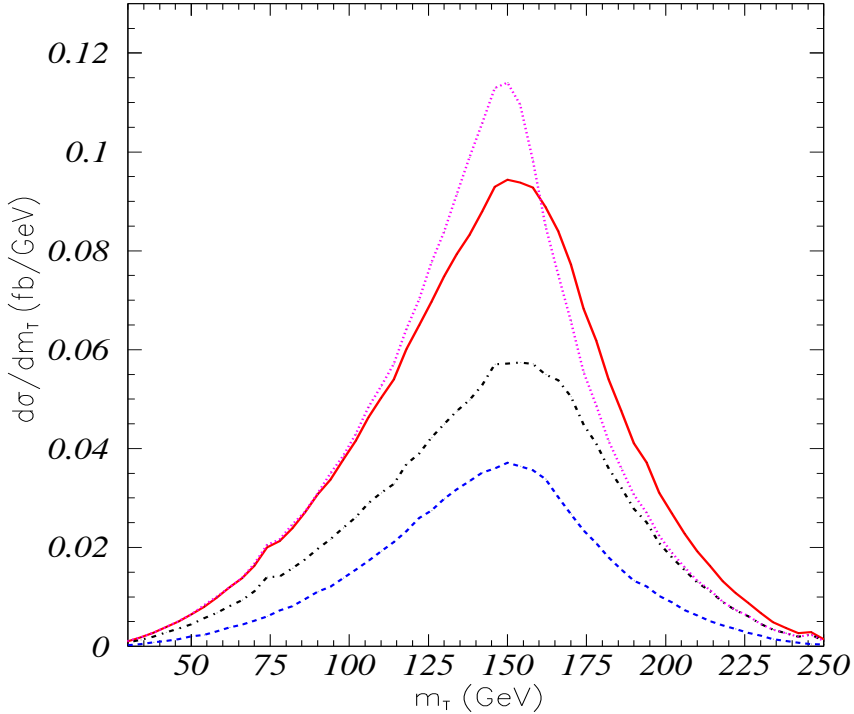,width=0.75\linewidth,height=0.65\linewidth}
\end{center}
\vspace{-2mm}
\caption{Transverse mass $m_T$ spectrum for the continuum model signal
  (dashed), background (dash-dotted), and continuum signal plus
  background (solid).  For comparison we also show the $m_h=155$~GeV
  SM signal plus background case, normalized to have the same signal
  cross section as the continuum model.}
\label{fig:sigback}
\end{figure}

For a continuum model we indeed observe a transverse mass peak around
the $WW$ threshold, but with a considerable tail extending to larger
$m_T$ values and therefore larger continuum Higgs masses. For a flat
continuum mass spectrum, the position of the peak in
Fig.~\ref{fig:sigback} is approximately the peak position of the $m_T$
curve given by a SM Higgs whose mass maximizes the function
$\sigma^{SM}_{Hjj}(m_h)\times B_{WW}(m_h)$.

\begin{table}[t]
\caption{Signal and background cross sections [fb] after all kinematic 
cuts as discussed in the text.  The first two columns refer to 
different-flavor ($e\mu$) final states, while the last two columns 
give results for the same-flavor ($ll=ee,\mu\mu$) sample.  The first
row is the continuum Higgs model considered, and we show the SM signal
which has the same peak value of $m_T(WW)$ in the second row for 
comparison.  The last three rows display the total background for each
channel, with and without the minijet veto; the signal-to-background
ratio for the continuum model; and the required integrated luminosity
to observe the continuum model signal over the SM background.}
\label{tab:channels}
\vspace{1.5mm}
\begin{tabular}{l||c|c||c|c}
channel &
$e^\pm \mu^\mp \phantom{ii}$ & 
$e^\pm \mu^\mp$ w/ minijet veto $\phantom{i}$ &
$e^\pm e^\mp$, $\mu^\pm\mu^\mp \phantom{i}$ &
$e^\pm e^\mp$, $\mu^\pm\mu^\mp$ w/ minijet veto \\ 
\hline 
$70<m_h<300\GeV$   & 1.90  & 1.69  & 1.56  & 1.39  \\ 
SM, $m_h=155\GeV$  & 5.60  & 4.98  & 4.45  & 3.96  \\
\hline 
$t\bar{t}$         & 0.086 & 0.025 & 0.086 & 0.025 \\ 
$t\bar{t}j$        & 7.59  & 2.20  & 6.45  & 1.87  \\ 
$t\bar{t}jj$       & 0.83  & 0.24  & 0.72  & 0.21  \\ 
single-top ($tbj$) & 0.020 & 0.015 & 0.016 & 0.012 \\
$b\bar{b}jj$       & 0.010 & 0.003 & 0.003 & 0.001 \\ 
QCD $WWjj$         & 0.448 & 0.130 & 0.390 & 0.113 \\ 
EW  $WWjj$         & 0.269 & 0.202 & 0.239 & 0.179 \\ 
QCD $\tau\tau jj$  & 0.128 & 0.037 & 0.114 & 0.033 \\ 
EW  $\tau\tau jj$  & 0.017 & 0.013 & 0.016 & 0.012 \\
QCD $\ell\ell jj$  &  --   &  --   & 0.114 & 0.033 \\ 
EW  $\ell\ell jj$  &  --   &  --   & 0.011 & 0.008 \\
\hline
total bkg          & 9.40  & 2.87  & 8.04  & 2.49  \\
$S/B$              & 1/5.0 & 1/1.7 & 1/5.1 & 1/1.8 \\
${\cal L}^{\rm obs}_{5\sigma} [\fb^{-1}]$ & 65 & 25 & 82 & 32 \\
\end{tabular}
\end{table}

We can see from Table~\ref{tab:channels} that all backgrounds are very
much under control, even with a reduced signal rate compared to the SM
Higgs.  In the intermediate mass range a SM Higgs could be observed
with $S/B=2\cdots 6$.  Now, with a wide mass window, this fraction is
reduced by roughly a factor $3$.  Because the luminosity required to
find a SM Higgs of mass $160\GeV$ at $5\sigma$ in this channel is less
that $5\fb^{-1}$, the luminosity required to discover a continuum
Higgs sector is still small.  For a flat distribution ($C(m_h)=C_0$)
over the range $70\leq m_h\leq 300\GeV$, $14\fb^{-1}$ are necessary to
observe a $5\sigma$ signal with a transverse mass peak as shown in
Fig.~\ref{fig:sigback}.

\subsection{Distinguishing Continuum from Single Resonance}
\label{sec:distSM}

While continuum Higgs sector discovery potential depends strongly on
the production rate and Higgs branching fraction to $W$ bosons,
$B_{WW}(m_h)$, these numbers depend strongly on the precise shape of
$C(m_h)$.  We propose to use the $WW$ transverse mass distribution
shape to distinguish SM from continuum Higgs sectors.  For reasons
discussed Sec.~\ref{sec:disc}, a continuum Higgs sector exhibits a
nearly symmetric distribution about the peak, with maximum around
$150\GeV$.  For small masses the rate is cut off by diminishing
branching fraction, and for high masses by the increasingly
restrictive production phase space (cf. Fig.~\ref{fig:1}).  In
contrast, a SM Higgs $m_T$ distribution is much more Jacobian.  After
determining the $m_T$ maximum we define the asymmetry
\begin{equation}
{\cal A} = \frac{ \sigma(m_T<m_T^{\rm max})
                 -\sigma(m_T>m_T^{\rm max})}
                { \sigma(m_T<m_T^{\rm max})
                 +\sigma(m_T>m_T^{\rm max})}
\equiv \frac{L-H}{L+H} = \frac{L-H}{N} \; ,
\label{eq:asymmetry}
\end{equation}
where $L(H)$ is the number of signal events with $m_T<(>)m_T^{\rm
  max}$, and $N$ is the total number of signal events.  We expect the
asymmetry to be small if the observed Higgs signal comes from a
continuum or multiple-fields model.

To quantify the impact of this asymmetry we again assume a flat
distribution, $C(m_h)=C_0$, over $70\leq m_h\leq 300\GeV$.  Assuming
that the total rate cannot serve as a means to distinguish the
continuum model from the SM Higgs sector, we normalize the $155\GeV$
Higgs SM rate to the continuum rate, after all cuts. This could \eg~be
realized by an overall suppression of the $HWW$ coupling, such as
occurs via $\sin(\beta-\alpha)$ in a two Higgs doublet model.  We then
evaluate the transverse $WW$ mass asymmetry ${\cal A}$ for signal and
background and compute the luminosity required to observe a continuum
signal five standard deviations away from the SM value.

For uncorrelated statistical errors $\sigma_l(\sigma_h)$ on $L(H)$,
the statistical significance of $\cal A$ is~\cite{Lyons:em}
\begin{equation}
\sigma_{\cal A} = 
\frac{2LH}{N^2} \sqrt{  \left(\frac{\sigma_L}{L}\right)^2
                      + \left(\frac{\sigma_H}{H}\right)^2 } =
\frac{1-{\cal A}^2}{2} \sqrt{ \frac{1}{L} + \frac{1}{H} } \; .
\end{equation}
where the last part of the relation holds if the number of events
satisfies Poisson statistics.

We show our results for several choices of continuum Higgs model
parameterizations in Table~\ref{tab:asymmetry}.  Any continuum sector
would be very easy to observe, typically requiring only around
10~fb$^{-1}$ to observe, although a case designed to be difficult
because of low branching ratio to $WW$, $70<m_h<150$~GeV, might
require up to 50~fb$^{-1}$ to observe at $5\sigma$.  Discriminating a
continuum sector from a SM Higgs presenting a peak in the same
position in $m_t$ is viable, but requires significant additional
integrated luminosity, typically ${\cal O}(100)$~fb$^{-1}$.  The
difficult cases, requiring close to ten times that amount of data, are
where the continuum is spread over very low masses, typically
$m_h<150$~GeV, such that very little of the total cross section decays
to $WW$ final states and the events lie below the nearly symmetric
peak in $\sigma^{SM}_{Hjj}(m_h)\times B_{WW}(m_h)$.

Note that we could have fitted the continuum $m_T$ line shape to a SM
Higgs mass, but we prefer to determine the SM mass to compare with
from the peak of the transverse mass distribution.  The only problem
we can think of is the mass distribution $C(m_h)$ mimicking the
transverse mass distribution of the SM Higgs boson.  However, this
only impacts the distinction from the SM and leaves the discovery
prospects unaffected.

\begin{table}[t]
\caption[]{The asymmetry defined in Eq.~(\ref{eq:asymmetry}) for SM 
Higgs signals and various choices of continuum Higgs sector.  We choose 
the SM Higgs mass of each pairing to give the same peak position in 
$m_T$ as the corresponding continuum model.  The first pair corresponds 
to the widest mass range; all choices assume $C(m_h)=C_0$.  The cross 
sections include all cuts and minijet veto survival probabilities. 
${\cal L}^{\rm obs}_{5 \sigma}$ is the luminosity required to observe 
the signal above background (``detectability'').  
${\cal L}^{\cal A}_{5 \sigma}$ is the luminosity required to 
distinguish the signal, using the $m_T$ asymmetry $\cal A$, as coming 
from a continuum model or a $150-170\GeV$ SM Higgs boson, depending on 
the case (``distinguishability'').}
\label{tab:asymmetry}
\vspace{1.5mm}
\begin{tabular}{ccccc}
  $\qquad$ Higgs mass (window) $\qquad$ 
& $\sigma$ [fb] & ${\cal L}^{\rm obs}_{5 \sigma}$ [fb$^{-1}$]
& ${\cal A}$    & ${\cal L}^{\cal  A}_{5 \sigma}$ [fb$^{-1}$] \\[1mm]
\hline
$\begin{matrix}{{\rm SM}: 155\GeV}\\{ 70\cdots 300\GeV}\end{matrix}$ &
$\begin{matrix}{ 14.3}\\{ 8.40}\end{matrix}$ &
$\begin{matrix}{   <5}\\{   14}\end{matrix}$ &
$\begin{matrix}{0.186}\\{0.076}\end{matrix}$ & 238 \\
\hline
$\begin{matrix}{{\rm SM}: 155\GeV}\\{100\cdots 270\GeV}\end{matrix}$ &
$\begin{matrix}{ 14.3}\\{ 9.40}\end{matrix}$ &
$\begin{matrix}{   <5}\\{    8}\end{matrix}$ &
$\begin{matrix}{0.210}\\{0.085}\end{matrix}$ & 161 \\
\hline
$\begin{matrix}{{\rm SM}: 155\GeV}\\{130\cdots 240\GeV}\end{matrix}$ &
$\begin{matrix}{ 14.3}\\{ 11.0}\end{matrix}$ &
$\begin{matrix}{   <5}\\{    4}\end{matrix}$ &
$\begin{matrix}{0.241}\\{0.081}\end{matrix}$ &  84 \\
\hline
$\begin{matrix}{{\rm SM}: 160\GeV}\\{185\cdots 300\GeV^{a}}\end{matrix}$ &
$\begin{matrix}{ 16.7}\\{ 7.37}\end{matrix}$ &
$\begin{matrix}{   <5}\\{   32}\end{matrix}$ &
$\begin{matrix}{0.231}\\{0.082}\end{matrix}$ & 145 \\
\hline
$\begin{matrix}{{\rm SM}: 170\GeV}\\{185\cdots 210\GeV}\end{matrix}$ &
$\begin{matrix}{ 16.2}\\{ 10.1}\end{matrix}$ &
$\begin{matrix}{   <5}\\{    6}\end{matrix}$ &
$\begin{matrix}{0.254}\\{0.125}\end{matrix}$ & 139 \\
\hline
$\begin{matrix}{{\rm SM}: 155\GeV}\\{ 70\cdots 180\GeV}\end{matrix}$ &
$\begin{matrix}{ 14.3}\\{ 9.43}\end{matrix}$ &
$\begin{matrix}{   <5}\\{    8}\end{matrix}$ &
$\begin{matrix}{0.211}\\{0.155}\end{matrix}$ & 810 \\
\hline
$\begin{matrix}{{\rm SM}: 150\GeV}\\{ 70\cdots 150\GeV}\end{matrix}$ &
$\begin{matrix}{ 12.6}\\{ 6.92}\end{matrix}$ &
$\begin{matrix}{   <5}\\{   52}\end{matrix}$ &
$\begin{matrix}{0.078}\\{0.141}\end{matrix}$ & 897 \\
\end{tabular}
{\small $^{a}$ Satisfies the slightly weaker constraint 
$\langle M^2 \rangle \, \lesssim (214 \GeV)^2$ 
(see Eq.~\ref{constraint}).}
\end{table}
%

\section{Remaining LHC Higgs Boson Searches}
\label{sec:others}

Having described how the multi Higgs resonance model is naturally
picked up by the WBF Higgs search with decay to $W$ bosons, we briefly
describe how the other LHC Higgs boson search channels would be
affected by this model.

\underline{Weak boson fusion:} The decay to $W$ pairs is not the only
decay channel proven capable of giving rapid Higgs boson discovery in
WBF processes. The decay to tau leptons in this production channel is
in fact the most solid channel in the MSSM, where it can even
distinguish between the two CP-even scalar mass
peaks~\cite{MSSMnolose,WBF.taus}. However, a mass peak is precisely
where the multi-resonance model avoids discovery. No matter how wide
the spread in Higgs boson masses becomes, the WBF tau decay channel
can see only the range between about 115 and 145~GeV. In that range
the signal would be approximately flat in tau pair invariant mass,
appearing to be an unexplained enhancement in the $Z\to\tau\tau$
Breit-Wigner tail above the peak. WBF production and decay to photon
pairs is also a useful channel~\cite{Rainwater:1997dg}, but similarly
would give an essentially flat invariant mass distribution. Because
this channel yields very few events, it works in the SM case only
because the detectors have a very narrow resolution in photon pair
invariant mass. Spreading a very small number of signal events out
from a 2~GeV window to over a $\sim 40$~GeV range would result in
complete loss of detectability. Seeing a Higgs signal in the WBF
channel decaying to a $W$ pair and not seeing it in the $\tau$ decays
in an accessible mass range would point to a Higgs sector beyond the
SM, possibly a continuum model.

\underline{Gluon fusion:} The two most powerful decay channels for
measuring the Higgs boson mass are photon pairs and the ``golden
mode'' of four leptons (via a $Z$ boson
pair)~\cite{atlas,cms,Cranmer:2003kf}, produced in gluon
fusion~\cite{gg.nlo}. Unfortunately, this feature turns into the worst
disadvantage once the mass window in the analysis has to be several
tens of GeV. A side-band analysis of the continuum background will not
work because there is no true resonance, and the significance of the
signal will vanish. On the other hand, we expect the $WW$ decay would
work similarly well as in the the WBF process. The $\ell\ell\nu\nu$
invariant mass shown in Fig.3 of Ref.~\cite{Dittmar:1996ss} will
change in analogy to the transverse mass in the WBF channel, but
without a detailed simulation (beyond the scope of this work) it is
not clear if the shape change would allow SM v. multi-resonance Higgs
sector discrimination.

\underline{Top quark associated production:} Despite the relatively
small cross section~\cite{ttH.nlo}, top quark events are highly
distinctive and associated production with a Higgs boson has as a
result received much interest.  For $m_h<135$~GeV the dominant decay
mode is to a pair of bottom quarks, which appears as a very small peak
near the $t\bar{t}b\bar{b}$ continuum peak~\cite{ttH.bb}.  Such a
signal spread out via continuum Higgs production would obviously be
immediately lost in the background.  Similarly for decay to photon
pairs, as in WBF or inclusive Higgs boson production. For
$m_h>135$~GeV the dominant decay is to $W$ pairs. This will be a very
useful channel for measuring the top Yukawa coupling~\cite{ttH.WW}, if
the Higgs sector is SM-like and $m_h>135$~GeV.  However, the
multi-lepton final states used in this planned analysis do not permit
full reconstruction, so the Higgs boson mass is never identified.
Because of the lack of need for such a resonance, this channel should
work nearly as well as in the SM case, although a detailed
investigation is again beyond the scope of our present work.

\underline{Weak boson associated production:} If the Higgs boson is
produced in association with a weak gauge boson it can be searched for
in the photon decay channel~\cite{atlas,cms}, because of the very
narrow resonance peak distinguishable above a large background.
However, just as in the case of decay to photon pairs in all the other
production channels, there is no longer any resonance, and the signal
will be lost in the continuum background. There is also a small signal
for $H\to b\bar{b}$ in the SM~\cite{WH.bb}, but as with $t\bar{t}H$
production the peak is similarly lost in the very large $Wb\bar{b}$
QCD continuum~\cite{Campbell:2003hd}. Thus, none of these modes would
be observable.

\section{Conclusions}

We have shown that continuum Higgs models do not present any problem
for LHC Higgs discovery.  If the coupling to gauge bosons is spread
over a mass range $m_h\sim 100\cdots 300\GeV$ these continuum Higgs
events will automatically appear in the searches for Higgs boson
produced in weak boson fusion, with subsequent decay to $\ell^+\ell^-$
and missing transverse energy via $W$ pairs.  The broad nature of the
Higgs resonance has no major impact on this search channel, because
the corresponding Standard Model search channel already benefits from
hugely suppressed backgrounds, typically much less than the level of
the signal.  In the case of a continuum Higgs sector, the signal to
background ratio typically is in the range $1/1-1/3$, still extremely
good.  Discovery of the most difficult case studied would require only
about 50~fb$^{-1}$, and almost all cases require less than
30~fb$^{-1}$.  Our analysis is not subject to detector uncertainties
such as poorly-modeled $\sla{E}_T$ resolution, or identification and
measurement of the tagging jets, that are known with less certainty
for high-luminosity running; only low-luminosity running is needed for
discovery in this channel.

Using the nearly symmetric behavior of the transverse $WW$ mass for
continuum Higgs sectors, we can distinguish these models from the SM
without relying on the total rate.  These features are largely
independent of the detailed spectrum $C(m_h)$ of the continuum Higgs
sector and of the distribution of the coupling $g_{WWh}(m_h)$.  The
integrated luminosity required to distinguish such a sector from the
SM is typically ${\cal O}(100)$~fb$^{-1}$, but can be as large as
900~fb$^{-1}$ for the most difficult case studied, where the continuum
is spread over a mass region less than 150~GeV.

Search strategies not using $WW$ final states will almost universally
be unable to see these states, as they require a sharp peak in a mass
spectrum, which does not exist in the flat-spectrum continuum models.
Such an observation could be interpreted as a sign of a Higgs sector
more complex than that of the SM, \eg~a continuum sector.


\begin{acknowledgments}
  
  This research was supported in part by Funda\c{c}\~{a}o de Amparo
  \`a Pesquisa do Estado de S\~ao Paulo (FAPESP), by Conselho Nacional
  de Desenvolvimento Cient\'{\i}fico e Tecnol\'ogico (CNPq), and by
  Programa de Apoio a N\'ucleos de Excel\^encia (PRONEX).

\end{acknowledgments}



\clearpage


\begin{thebibliography}{0}

\bibitem{Group:2002mc}
D.~Abbaneo {\it et al.}  [LEPEWWG],
arXiv:hep-ex/0212036.

\bibitem{Dittmar:1996ss}
M.~Dittmar and H.~K.~Dreiner,
 Phys.\ Rev.\ D {\bf 55}, 167 (1997).

\bibitem{Rainwater:1999sd}
D.~Rainwater and D.~Zeppenfeld,
 Phys.\ Rev.\ D {\bf 60}, 113004 (1999) \\
 {[Erratum-ibid.\ D {\bf 61}, 099901 (2000)]}.

\bibitem{Kauer:2000hi}
N.~Kauer, T.~Plehn, D.~Rainwater and D.~Zeppenfeld,
 Phys.\ Lett.\ B {\bf 503}, 113 (2001).

\bibitem{WBF.taus}
D.~Rainwater, D.~Zeppenfeld and K.~Hagiwara,
Phys.\ Rev.\ D {\bf 59}, 014037 (1999); \\
T.~Plehn, D.~Rainwater and D.~Zeppenfeld,
Phys.\ Rev.\ D {\bf 61}, 093005 (2000).

\bibitem{WBF_exp1}
C.~Buttar, R.~Harper, K.~Jakobs, ATL-PHYS-2002-033.

\bibitem{WBF_exp2}
S.~Asai {\it et al.}, ATL-PHYS-2003-005; 
B.~Mellado, ATL-CONF-2002-004; \\
N.~Akchurin {\it et al.}, CMS-NOTE-2002/066; \\
G.~Azuelos and R.~Mazini, ATL-PHYS-2003-004; \\
K.~Cranmer {\it et al.}, ATL-PHYS-2003-008 and ATL-PHYS-2003-007.

\bibitem{atlas}
ATLAS TDR, report CERN/LHCC/99-15 (1999).

\bibitem{cms}
CMS TP, report CERN/LHCC/94-38 (1994).

\bibitem{MSSMnolose}
T.~Plehn, D.~Rainwater and D.~Zeppenfeld,
Phys.\ Lett.\ B {\bf 454}, 297 (1999).

\bibitem{Ellwanger:2003jt}
U.~Ellwanger, J.~F.~Gunion, C.~Hugonie and S.~Moretti,
arXiv:hep-ph/0305109.

\bibitem{stealth}
T.~Binoth and J.~J.~van der Bij,
arXiv:hep-ph/9409332; 
Z.\ Phys.\ C {\bf 75}, 17 (1997).

\bibitem{invis}
O.~J.~P.~\'Eboli and D.~Zeppenfeld,
Phys.\ Lett.\ B {\bf 495}, 147 (2000).

\bibitem{Espinosa:1998xj}
J.~R.~Espinosa and J.~F.~Gunion,
Phys.\ Rev.\ Lett.\  {\bf 82}, 1084 (1999).

\bibitem{Gunion:2001pg}
J.~F.~Gunion,
arXiv:hep-ph/0106154.

\bibitem{Espinosa:1998re}
J.~R.~Espinosa and M.~Quiros,
Phys.\ Rev.\ Lett.\  {\bf 81}, 516 (1998).

\bibitem{:2003sz}
[LEP Higgs Working for Higgs boson searches Collaboration],
arXiv:hep-ex/0306033.

\bibitem{Pumplin:2002vw}
J.~Pumplin et al.
JHEP {\bf 0207}, 012 (2002).

\bibitem{Stelzer:1994ta}
T.~Stelzer and W.~F.~Long,
Comput.\ Phys.\ Commun.\  {\bf 81}, 357 (1994).

\bibitem{OS-tops}
N.~Kauer,
Phys.\ Rev.\ D {\bf 67}, 054013 (2003); \\
N.~Kauer and D.~Zeppenfeld,
Phys.\ Rev.\ D {\bf 65}, 014021 (2002).

\bibitem{Willenbrock:cr}
S.~S.~D.~Willenbrock and D.~A.~Dicus,
Phys.\ Rev.\ D {\bf 34}, 155 (1986).

\bibitem{Ellis:1987xu}
R.~K.~Ellis, I.~Hinchliffe, M.~Soldate and J.~J.~van der Bij,
Nucl.\ Phys.\ B {\bf 297}, 221 (1988).

\bibitem{Lyons:em}
L.~Lyons, {\underline Statistics for Nuclear and Particle Physicists},
Cambridge University Press, 1986.

\bibitem{Rainwater:1997dg}
D.~Rainwater and D.~Zeppenfeld,
JHEP {\bf 9712}, 005 (1997).

\bibitem{Cranmer:2003kf}
K.~Cranmer, B.~Mellado, W.~Quayle and S.~L.~Wu,
ATL-PHYS-2003-025.

\bibitem{gg.nlo}
 M.~Spira, A.~Djouadi, D.~Graudenz and P.~M.~Zerwas,
  Nucl.\ Phys.\ B {\bf 453}, 17 (1995); \\
 R.~V.~Harlander and W.~B.~Kilgore,
  Phys.\ Rev.\ Lett.\  {\bf 88}, 201801 (2002); \\
 C.~Anastasiou and K.~Melnikov,
  Nucl.\ Phys.\ B {\bf 646}, 220 (2002); \\
 V.~Ravindran, J.~Smith and W.~L.~van Neerven,
  Nucl.\ Phys.\ B {\bf 665}, 325 (2003); Mod.\ Phys.\ Lett.\ A {\bf 18}, 1721 (2003).

\bibitem{ttH.nlo}
W.~Beenakker {\it et al.},
Phys.\ Rev.\ Lett.\  {\bf 87}, 201805 (2001); \\
S.~Dawson, L.~H.~Orr, L.~Reina and D.~Wackeroth,
Phys.\ Rev.\ D {\bf 67}, 071503 (2003).

\bibitem{ttH.bb}
V.~Drollinger, T.~M\"uller and D.~Denegri,
arXiv:hep-ph/0111312; \\
J.~Cammin and M.~Schumacher, ATL-PHYS-2003-024.

\bibitem{ttH.WW}
F.~Maltoni, D.~Rainwater and S.~Willenbrock,
Phys.\ Rev.\ D {\bf 66}, 034022 (2002); \\
V.~Kostioukhine, J.~Leveque, A.~Rozanov, J.~B.~de~Vivie,
ATL-PHYS-2002-019.

\bibitem{WH.bb}
V.~Drollinger, T.~M\"uller and D.~Denegri,
arXiv:hep-ph/0201249; \\
B.~P.~Kersevan and E.~Richter-Was,
Eur.\ Phys.\ J.\ C {\bf 25}, 379 (2002).

\bibitem{Campbell:2003hd}
J.~Campbell, R.~K.~Ellis and D.~Rainwater,
arXiv:hep-ph/0308195.

\end{thebibliography}
\end{document}